\documentstyle[twocolumn,epsf,prl,aps]{revtex}

\begin{document}
\draft

\title{Sand as Maxwell's demon}

\author{
Jens Eggers
}

\address{
Universit\"{a}t Gesamthochschule Essen, Fachbereich Physik,
45117 Essen, Germany  }

\maketitle
\begin{abstract}
We consider a dilute gas of granular material inside a box,
kept in a stationary state by shaking. A wall separates 
the box into two identical compartments, save for a small hole 
at some finite height $h$. As the gas is cooled, a second 
order phase transition occurs, in which the particles preferentially
occupy one side of the box. We develop a quantitative theory
of this clustering phenomenon and find good agreement with 
numerical simulations. 
\end{abstract}


One of the most outstanding features of a gas of granular material
is its tendency to spontaneously form highly concentrated regions
or clusters \cite{JaNa92,GZ93,MY92}. So even in its gaseous state 
it behaves fundamentally different from a molecular gas, which keeps
its uniform density. Apart from throwing light on the nonequilibrium
properties of a granular gas, understanding this clustering instability 
is of major technological importance. Imagine a flow of rocks down
a chute: whenever a very dense region has formed due to the instability,
the rocks may easily get entangled and the flow is stuck. 

The clustering instability is caused by the 
distinguishing feature of a granular gas
in contrast to a molecular gas, namely that a fraction of the kinetic energy 
is irretrievably lost upon a collision, and simply heats the particles. 
In this paper, we will concentrate on the simplest case of a dilute
gas for which collisions are binary, so the rate of energy 
loss grows quadratically with the density. This means a dense region
rapidly cools, increasing the density even more according to the 
equation of state. Goldhirsch and Zanetti \cite{GZ93} used a hydrodynamic
description of a dilute gas or ``rapid granular flow'' 
\cite{JeSa83,JR85,JeRi86,C90,SeGo98} to show that this mechanism
leads to a long-wavelength instability of a homogeneous assembly 
of inelastic particles. However, in the absence of driving, the collapse 
quickly becomes complete and the original assumptions break down.
In the system described in this letter, a stationary state is 
maintained by external driving. It is thus simple enough to
allow for an analytical description, while it keeps the central 
aspects of the clustering phenomenon. 

\begin{figure}
\begin{center}
\leavevmode
\epsfsize=0.6 \textwidth
\epsffile{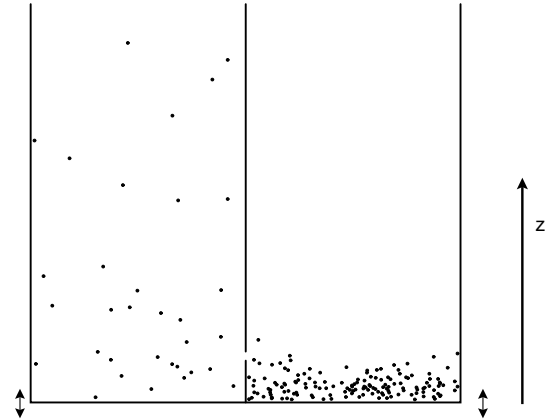}
\end{center}
\caption{
A drawing of the experimental setup. The two sides of the 
box are connected by a hole at height $h$. The picture is
taken below the bifurcation, so most particles have settled
on the right hand side. As a result, the gas sinks to the bottom,
reducing the flux.
}
\label{box}
\end{figure}

The experiment, first described by Schlichting and Nordmeier \cite{SN96},
consists of a box of base area $12 cm^2$ and height $20 cm$, mounted on
a shaker, and filled with $N = 100$ plastic particles of radius 
$r = 1mm$ (see Fig. \ref{box}). The box is separated into two equal parts 
by a wall which has a narrow horizontal slit at a height $h = 2.3cm$. 
When the shaker is operating at full power, the amplitude is vibration is 
approximately $A = 0.3 cm$ and the frequency $f = 50 Hz$. Even
if all particles are on one side initially, they immediately distribute
equally to both sides. Lowering the frequency below a critical value 
of $30 Hz$ the symmetry is spontaneously broken, and particles settle
preferentially on one side. Evidently, the wall greatly enhances 
the clustering described above, since it prevents a direct exchange
of particles which would break up the cluster. The grains thus act
as Maxwell's demon \cite{LR90}, who preferentially lets particles 
pass from left to right or vice versa. As a result, a more ordered state
is formed in which most particles are on one side. The demon must
then absorb entropy, a role which in our system is assumed by the 
sand grains. Still another interpretation would be that of a dissipative 
structure, which in the stationary state is maintained by a flux of 
entropy \cite{GP}.

The condition of stationarity is that the total flux between the two 
compartments is zero, i.e. the fluxes going from one side to the other
cancel:
\begin{equation}
\label{stat}
F_{\ell\rightarrow r} (h) = F_{r\rightarrow \ell}(h) .
\end{equation}
In the limit that the connecting hole is small, it remains to find the density 
and temperature distribution in a one-dimensional column of a vibrated 
granular assembly, from which we calculate the flux of particles leaving a 
compartment at the height of the hole $h$. Below we show that (\ref{stat})
has a single solution for strong driving, but two asymmetric 
solutions become stable as the driving is lowered. 

The problem of a shaken granular gas in a gravitational field 
has recently been treated in a number of papers \cite{L95,K98,ML98}.
Equations of motion for the number density $n(z)$, pressure $p(z)$,
and granular temperature $T(z)$ can be found from conventional 
kinetic theory \cite{JeSa83,JR85}. The position $z$ is measured 
from the bottom of the container and the granular temperature 
is defined as $T = <v^2>/d$, where $d$ is the dimension of space
and $v$ the velocity of a particle. This definition of temperature 
is customary for granular media, but differs from the usual 
molecular temperature, which is recovered by formally setting $k_B = m$.
Now the stationary equations become in the dilute limit
\begin{eqnarray}
\label{equ}
&&p = m n T,\quad  \partial_z p = -mg n , \nonumber \\
&&\kappa\partial_z \left[T^{1/2}\partial_z T\right] =
(2\kappa/3)\partial^2_z T^{3/2} = D n^2 T^{3/2} .
\end{eqnarray}
The first equation is the usual equation of state, the second
the force balance, where $-mg$ is the force on a single 
particle. The third equation is the balance of heat flux and
dissipation due to inelastic collisions. As usual, the thermal 
conductivity $\kappa T^{1/2}$ is proportional to the average 
particle velocity. In the simplest possible model of hard and
smooth particles the energy loss in one collision is proportional 
to $(1-e^2) T$ on the average. The coefficient of restitution 
$e$ measures the reduction $v'_n = -e v_n$ in the normal velocity 
of the particles. Together with the number of collisions being
proportional to $n^2 T^{1/2}$ this explains the form of the loss term 
on the right hand side of the third equation (\ref{equ}).

To allow for a better comparison with numerical simulations, 
we will consider the two-dimensional case of circular discs 
of radius $r$, for which the coefficients are found to be \cite{JR85}
$\kappa = \pi^{-1/2} m/r, \quad D = 4 \pi^{1/2} mr(1-e)$.
Next, we have to supply boundary conditions. To minimize wall 
effects, which are not essential to our problem, all wall collisions 
are assumed to be elastic. The top is left open.
For simplicity, the bottom of the 
container is taken to move in a sawtooth manner, such that 
a colliding particle always finds it to move upward with velocity 
$v_b = Af$. Finally, the amplitude $A$ of the vibration is assumed to be 
very small compared with the mean free path, 
so that the bottom is effectively stationary. 
In summary this means that the z-component of the velocity of
a particle colliding with the bottom is changed according to 
$v'_z = 2v_b - v_z$. The approximations of (\ref{equ}) imply
that the velocity is close to a Maxwell-Boltzmann distribution,
which allows to calculate the rate of energy input per unit 
width (or unit area in 3D) to 
\begin{equation}
\label{Q}
Q = mn\left[v_b T(0) + \sqrt{2/\pi} v_b^2 T(0)^{1/2}\right] .
\end{equation}
We will see below that the first term in (\ref{Q}) dominates,
the second typically being smaller by a factor of 10 in our 
simulations. Thus the two boundary conditions for $z=0$ become 
\begin{equation}
\label{bound}
p(0) = -gm \bar{N},\quad Q = -\kappa T^{1/2} \partial_z T (0) .
\end{equation}
The first equation comes from integrating the force balance, 
and $\bar{N}$ is the total number of particles per unit width
of the box. The second equation balances the energy input with 
the heat flux out of the bottom according to (\ref{equ}).

\begin{figure}
\begin{center}
\leavevmode
\epsfsize=0.4 \textwidth
\epsffile{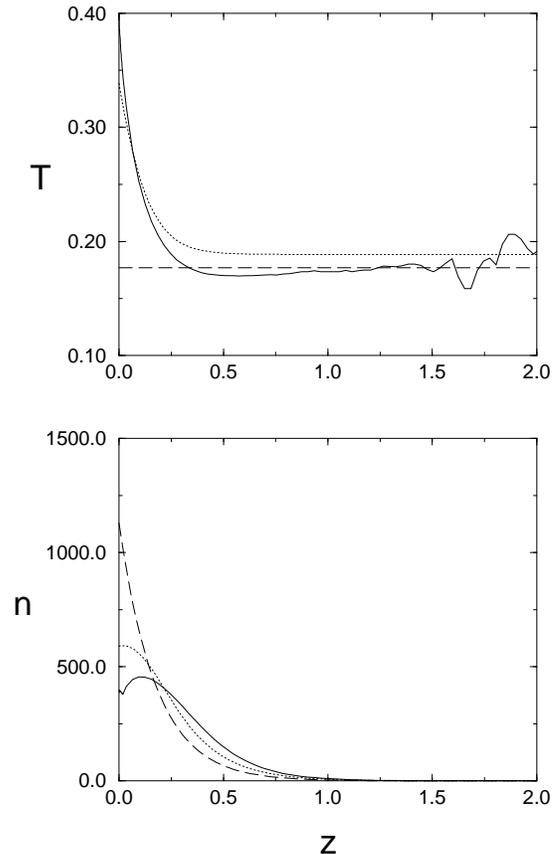}
\end{center}
\caption{
Temperature and density profiles of $N = 320$ grains 
in a box of width $w = 1.6$. The particles are circular 
disks of radius $r = 0.01$ and coefficient of restitution 
$e = 0.95$. The velocity of the bottom is $v_b = 0.149$,
acceleration of gravity $g$ and mass $m$ is normalized to one
The full line is the result of a particle simulation, the 
dotted line is the present theory, and the dashed line is
(\protect\ref{const}).
}
\label{prof}
\end{figure}

Since (\ref{equ}) is of third order, another boundary condition 
is needed for a unique solution. The missing third condition is
found by observing that the solution of (\ref{equ}) for large
$z$ is of the form $T^{3/2}(z) = cz + b$, so $c$ must be zero 
to prevent $T$ from becoming negative or infinity. This is the 
desired third condition, and solutions are easily found by shooting for a 
temperature profile which is asymptotically constant. The 
resulting temperature and density profiles are shown as dotted lines in 
Fig. \ref{prof} for a typical set of parameters and are compared 
with a numerical simulation (solid lines). The simulation uses an event-driven 
code \cite{R80} for smooth and hard particles, with the normal
velocity reduced by a factor of $e$ as described above. The agreement 
is quite good, but gets worse if the number of particles is reduced,
and deteriorates even more in three dimensions. The reason is that 
even for the parameters of Fig. \ref{prof} the temperature changes
significantly over the length of the mean free path. As a result, 
the distribution of the $v_z$-velocities of particles hitting the 
bottom deviates significantly from a Maxwell-Boltzmann distribution 
at $T(0)$. 

The solution given in \cite{K98} assumes that the temperature is
constant, which is asymptotically correct for very small inelasticity.
The resulting profiles are the dashed lines in Fig. \ref{prof}.
Evidently, the present theory is more accurate, but the constant 
temperature solution has the great merit of simplicity and still 
contains the essentials. Namely, at constant temperature the density 
is an exponential and 
$T_{\infty}$ is found from a balance of the energy input and dissipation:
\begin{equation}
\label{const}
n(z) = \frac{gm\bar{N}}{T_{\infty}} e^{-gz/T_{\infty}} ,\quad
T_{\infty} = \left[\frac{2v_b}{D m \bar{N}}\right]^2 .
\end{equation}
Here we have neglected the second term of (\ref{Q}) in favor
of the first, which is again consistent for $e \approx 1$.

Returning to the original stability problem, the {\it flux} through a 
hole of area $S$ is found to be 
$F = S n(h) \sqrt{T(h)/2\pi}. $
Using this formula, and solutions of (\ref{equ})-(\ref{bound})
to find the profiles, one can look for solutions of (\ref{stat}).
Since the total number of particles $N$ is conserved, these must 
be sought subject to the constraint $\bar{N} = \bar{N}_{\ell} +
\bar{N}_r$, where the overbar denotes the number of particles 
relative to the half-width of the container. 

\begin{figure}
\begin{center}
\leavevmode
\epsfsize=0.4 \textwidth
\epsffile{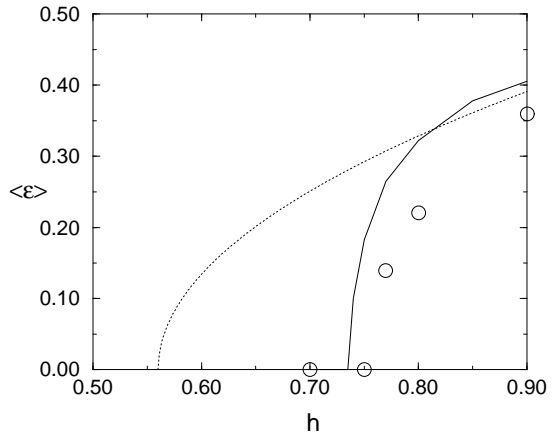}
\end{center}
\caption{
The bifurcation of the asymmetry $\epsilon$ as function 
of the height $h$. The full line is the result of the present 
theory, circles are numerical simulations with $N = 360$ 
particles in a box of half-width 1.6, averaged over time.
All other parameters 
are the same as in Fig. \protect{\ref{prof}}. The dotted line 
is the result (\protect\ref{mu}) of the simplified theory.
}
\label{bif}
\end{figure}

\begin{figure}
\begin{center}
\leavevmode
\epsfsize=0.4 \textwidth
\epsffile{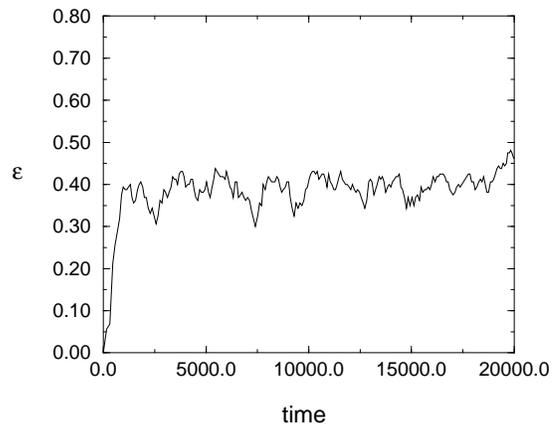}
\end{center}
\caption{
The temporal evolution of the asymmetry parameter $\epsilon$
corresponding to the data point $h = 0.9$ in Fig.\protect\ref{box},
but with only $N = 180$ particles.
}
\label{eps}
\end{figure}

Figure \ref{bif} shows the solutions of (\ref{stat}) as a function of 
$h$ for $\bar{N} = 225$ as the solid line. The 
instability follows a typical pitchfork bifurcation, so the average of the
asymmetry parameter $\epsilon = (\bar{N}_{\ell/r} - \bar{N})/\bar{N}$
increases continuously when $h$ is raised above a critical value. 
The circles, which are the results of a numerical simulation with 
$N = 360$ particles averaged over time, show quite satisfactory agreement. 
Figure \ref{eps} illustrates the temporal behavior of the asymmetry 
parameter $\epsilon$. The symmetric state rapidly becomes unstable 
and the system fluctuates around its new equilibrium position. 
Since there are more collisions on the denser side the temperature 
is lowered, causing particles to sink to the bottom.
Thus a non-symmetric state becomes possible because the flux
$F$ is no longer a monotonously increasing function of the number of 
particles as it would be in equilibrium. Another hallmark
of a nonequilibrium system is that the stationary state is described
by a flux balance (\ref{stat}), while the temperature 
on either side of the hole is not equal, as it would be in thermal 
equilibrium. 

Further insight can be gained from the simplified solution 
(\ref{const}). A straightforward calculation 
leads to 
\begin{eqnarray} 
\label{dte}
&&\partial_t\epsilon = F_0 N \nonumber \\
&&\left[(\epsilon-1/2)^2e^{-\mu (\epsilon-1/2)^2} - 
(\epsilon+1/2)^2e^{-\mu (\epsilon+1/2)^2} \right] + \xi,
\end{eqnarray}
where $\xi$ is a noise term to be considered later and 
$F_0$ is a constant. The stationary solutions for small noise are determined 
by the vanishing of the angular brackets, which is controlled by a single 
parameter 
\begin{equation}
\label{mu}
\mu = 4\pi ghr^2 (1-e)^2 \bar{N}^2 / v_b^2 .
\end{equation}
If $\mu > 4$, the equation $[\dots] = 0$ has three roots, 
and just above the bifurcation the asymmetric solutions are 
described by $<\epsilon> = \pm\sqrt{3(\mu-4)/16}$,
which is included as the dashed line in Fig. \ref{bif}.
While there is an offset between the simplified theory and
the simulation, it does describe the form of the bifurcation
fairly well using the single parameter $\mu$. It is intuitively
clear that by increasing the number of particles, the inelasticity, 
or the height of the wall separating the two compartments, 
phase separation is enhanced. Heating, as pointed out in the 
introduction, favors a symmetric state. 

Finally, it is possible to include fluctuations in our description, which 
are important if the number of particles is just a few hundred 
as in Fig. \ref{eps}. In particular this leads to a softening of the 
transition, since immediately below the transition the system 
can switch between the two possible states. Assuming that the particles 
passing through the hole are uncorrelated, the noise $\xi$ in 
(\ref{dte}) is Gaussian white noise \cite{R84} on a coarse-grained
time scale, and
\begin{eqnarray}
\label{xi} 
&&<\xi(t)\xi(t')> = F_0 \nonumber \\
&&\left[(\epsilon-1/2)^2e^{-\mu (\epsilon-1/2)^2} +
(\epsilon+1/2)^2e^{-\mu (\epsilon+1/2)^2} \right] \delta(t-t') .
\end{eqnarray}
Note that the constant $F_0$ can be eliminated by rescaling time, 
so the strength of the noise is controlled by the total number
of particles alone. By considering the fluctuations around the 
local minimum to Gaussian order, we find \cite{R84}
$\sqrt{<(\epsilon-<\epsilon>)^2>} = [4N(\mu-4)]^{-1/2}$.
Adjusting this formula to the critical value of $\mu_{cr} = 5.4$
found from simulation this gives $\sqrt{<(\epsilon-<\epsilon>)^2>} = 
0.045$ in reasonable agreement with Fig. \ref{eps}. Of course,
equations (\ref{dte})-(\ref{xi}) also allow for a more detailed 
analysis of transitions between the two asymmetric states and many
more questions relating to the critical fluctuations near the
transition. This will be considered in more detail in future 
publications. 

In conclusion, hydrodynamic equations are a very valuable 
tool to describe a dilute granular gas. The greatest problem 
lies in the formulation of the boundary conditions. The clustering
instability can be understood in terms of a very simple experiment,
which we model in a static, one-dimensional description.

I am very grateful to V. Nordmeier for showing me the experiment
this letter is based on. 
I have benefited greatly from conversations with J. Krug, S. Luding, 
W. Strunz, and J. Vollmer. J. Vollmer and S. Luding were tireless 
in their support. 
This work was also supported by the Deutsche Forschungsgemeinschaft
through SFB237.

\end{document}